\documentclass[aps,prl,twocolumn,showpacs,groupedaddress,preprintnumbers]{revtex4}

\usepackage{graphicx}
\usepackage{amsmath}
\usepackage[greek,english]{babel}
\usepackage{color}

\begin{document}

\hyphenation{re-com-men-ded fun-ction ope-ra-ting}

\preprint{\textbf{Submitted to Physical Review Letters on 21
December 2007. Resubmitted in revised form on 20 February 2008.}}

\title{Primary gas thermometry by means of laser absorption spectroscopy:\\Determination of the Boltzmann constant}

\author{G.~Casa}
\author{A.~Castrillo}
\author{G.~Galzerano}
\altaffiliation{Dipartimento di Fisica, Politecnico di Milano and
Istituto di Fotonica e Nanotecnologie (IFN-CNR), Milano, Italy.}
\author{R.~Wehr}
\author{A.~Merlone}
\altaffiliation{Istituto Nazionale di Ricerca Metrologica, Torino,
Italy.}
\author{D.~Di Serafino}
\altaffiliation{Dipartimento di Matematica, Seconda Universit\`{a}
di Napoli, Caserta, Italy.}
\author{P.~Laporta}
\altaffiliation{Dipartimento di Fisica, Politecnico di Milano and
Istituto di Fotonica e Nanotecnologie (IFN-CNR), Milano, Italy.}
\author{L.~Gianfrani}
\email{livio.gianfrani@unina2.it} \affiliation{Dipartimento di
Scienze Ambientali, Seconda Universit\`{a} di Napoli, Via Vivaldi
43, I-81100 Caserta, Italy}


\begin{abstract}
We report on a new optical implementation of primary gas
thermometry based on laser absorption spectrometry in the near
infrared. The method consists in retrieving the Doppler broadening
from highly accurate observations of the line shape of the R(12)
$\nu_{1}~+~2~\nu_{2}^{\phantom{1}0}~+~\nu_{3}$ transition in
CO$_{2}$ gas at thermodynamic equilibrium. Doppler width
measurements as a function of gas temperature, ranging between the
triple point of water and the gallium melting point, allowed for a
spectroscopic determination of the Boltzmann constant with a
relative accuracy of $\sim$1.6$\times$10$^{-4}$.
\end{abstract}


\maketitle


Atomic and molecular spectroscopy have played a major role in
fundamental metrology since the advent of highly monochromatic,
tunable laser sources. Spectroscopic experiments with a
metrological goal mostly consist in precision measurements of
absolute frequencies, such as the center frequency of the hydrogen
1s-2s line \cite{Niering-2000-4,Schwob-1999-4} and the strontium intercombination line
\cite{Boyd-2007-4}, just to mention a few significant examples.

Recently, high resolution laser absorption spectroscopy has been
proposed for a direct determination of the Boltzmann constant
\cite{borde-2005-4}. In this case, a frequency interval has to be
measured very accurately. In fact, the method relies on a
determination of the line width of a Doppler-broadened absorption
profile for an atomic or molecular line in a rarefied gas at
thermodynamic equilibrium. The current value of $k_{\mathrm{B}}$,
1.3806504~(24)~$\times$~10$^{-23}$~JK$^{-1}$, recommended by the
Committee on Data for Science and Technology (CODATA)
\cite{codata}, comes from the ratio between the molar gas constant
$R$ and the Avogadro number $N_{\mathrm{A}}$ \cite{mohr-2005-4}.
Its relative uncertainty, 1.7~$\times$~10$^{-6}$, is mostly due to
the uncertainty on $R$, whose accepted value is essentially given
by that obtained in 1988 by Moldover \textit{et al.}
\cite{Moldover-1998-4} through a fascinating experiment based on
acoustic gas thermometry in argon. There is presently a strong
interest in new and more direct determinations of $k_{\mathrm{B}}$
in view of a possible new definition of the unit kelvin
\cite{Fellmuth-2006-4}. In the current International System of
Units (SI), the unit kelvin is linked to an artefact, which is far
from being invariant in space and time. As already done for the
unit metre by fixing the speed of light $c$, the unit kelvin could
be redefined through the thermal energy by fixing the Boltzmann
constant. This new approach would make the definition more general
and independent of any material substance, technique of
implementation, and temperature. Nonetheless, before fixing its
value, $k_{\mathrm{B}}$ must be determined with an improved
uncertainty, possibly using different methods. Presently, the most
accurate methods to measure $k_{\mathrm{B}}$ are based on acoustic
gas thermometry, exploiting the relation between the speed of
sound in a gas and the thermodynamic temperature, and on
dielectric constant gas thermometry, based on the virial expansion
of the Clausius-Mossotti equation and applicable only to helium,
for which QED provides the static electric dipole polarizability
\cite{Cencek-2001-4}. Nonetheless, the experimental investigation
of new primary thermometric methods, based on radically different
principles, is strongly encouraged by the Consultative Committee
for Thermometry of the International Committee for Weights and
Measures, as reported in its Recommendation T~2 of 2005.

Doppler-broadened laser absorption spectroscopy can be implemented
in such a way as to act as a primary thermometric method, with the
advantage of being conceptually simple, applicable to any gas at
any temperature, in whatever spectral region. In contrast to other
methods also based on electromagnetic radiation measurements, such
as total radiation thermometry \cite{Martin-1998-4}, this new
approach does not require absolute radiation determinations. It is
well known that the Doppler width, $\Delta\nu_{D}$ (FWHM), of a
line (with a center frequency $\nu$) in an absorbing molecular gas
at thermodynamic equilibrium depends on the temperature $T$
through the equation:
\begin{equation}
\Delta\nu_{D} =
2\frac{\nu}{c}\sqrt{2\ln2k_{\mathrm{B}}\frac{T}{m}}
\label{eq:4:one}
\end{equation}
\\
where $m$ is the mass of the molecule.

The first spectroscopic determination of $k_{\mathrm{B}}$ has been
recently performed in the mid-infrared by the Daussy \textit{et
al.} on the $\nu_{2}$ asQ(6,3) rovibrational line of the ammonia
molecule $^{14}$NH$_{3}$ at a frequency of 28 953 694 MHz
\cite{Daussy-2007-4}. The absorption profile was observed with a
maximum signal-to-noise ratio (S/N) of 10$^{3}$, in the pressure
range between 1 and 10 Pa, at a temperature of 273.15 K, using a
CO$_{2}$ laser frequency stabilized on a OsO$_{4}$ line. Under
these low-pressure conditions, which are close to the Doppler
limit, by measuring the width of the absorption line as a function
of the pressure and extrapolating to zero pressure, it was
possible to deduce the Doppler width, which yielded an evaluation
of the Boltzmann constant with a relative uncertainty of
1.9$\times$10$^{-4}$ \cite{Daussy-2007-4}. This approach allows
for a very simple spectral analysis but requires an accurate
determination of the absolute pressures.

Here we report on a radically different implementation of laser
absorption spectroscopy for primary gas thermometry. In
particular, we demonstrate that it is possible to retrieve the gas
temperature from a molecular absorption profile even when the gas
pressure is sufficiently high that the line shape is far from the
Doppler limit, but sufficiently small that one can neglect the
averaging effect of velocity-changing collisions, so that the line
shape is given by the exponential of a Voigt convolution. We
perform absorption spectroscopy in the near-infrared on a CO$_{2}$
gas sample at thermodynamic equilibrium using a distributed
feed-back (DFB) diode laser, probing the R(12) component of the
$\nu_{1}+2\nu_{2}^{\phantom{1}0}+\nu_{3}$ combination band. In
contrast to the NH$_{3}$ molecule, our molecular target does not
exhibit any hyperfine structure.


\begin{figure}
\centering
\includegraphics[bb=200 240 580 425, scale=0.63]{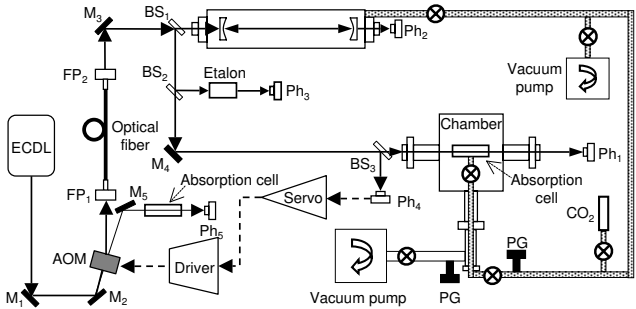}
\caption{\label{fig:4:one}  A sketch of the intensity-stabilized
laser absorption spectrometer. ECDL: extended-cavity diode laser;
M: mirror; BS: beam splitter; FP: fiber port; Ph: photodetector;
PG: pressure gauge. The cell is closed by an oil-free solenoid
valve, placed inside the chamber.}
\end{figure}

The experimental setup is depicted in Fig.~\ref{fig:4:one}. The
diode laser, emitting on a single mode at 2.006 $\mu$m with a
maximum power of 3 mW, is mounted in a mirror extended-cavity
configuration. In this scheme, a partially reflecting mirror, with
a 50\% reflection coating on one side and an antireflection (AR)
coating on the other, provides the optical feedback for line
narrowing down to $\sim$1 MHz. As will be shown later on, this is
sufficient for the aim of the present experiment. Collimated by a
double-aspheric lens, the laser beam first passes through an
acousto-optic modulator (AOM) and then is injected into a
single-mode, polarization-maintaining optical fiber. The latter is
employed as a high-quality spatial filter to produce an optimum
TEM$_{00}$ transverse mode. The AOM is used as an actuator within
a servo loop in order to keep the intensity of the laser beam
constant over a laser frequency scan as wide as 1 wavenumber.
Driven by a 1 W radiofrequency signal at 80 MHz, the AOM deflects
about 70\% of the laser power from the primary beam to the first
diffracted order, which is used for the experiment. The intensity
control feedback loop ensures a stability level better than
10$^{-4}$ (measured value over a laser frequency span of about 0.5
cm$^{-1}$). The servo loop is active within an electronic
bandwidth of 50 kHz. After being collimated again, the laser beam
is divided into four parts by a series of beam splitters. The
first beam is sent to a high-finesse, confocal optical resonator
that provides accurate frequency markers for calibration purposes.
The second goes to a 5 cm tick germanium etalon, which allows for
a coarse check of the frequency calibration. The third beam
impinges on a monitor photodiode, whose output signal is used as
the input signal to our servo to control the rf power driving the
AOM. Finally, the remaining portion of the laser beam passes
through a homemade, temperature-stabilized sample cell, closed at
each end by an AR-coated BK7 window. Temperature-stabilized,
extended-wavelength, InGaAs photodetectors are employed to monitor
the laser radiation. The optical resonator consists of a pair of
identical, high-quality, dielectric mirrors, with a radius of
curvature of 50 cm and a reflectivity of 99.5\%. The mirror
mounts, tightly locked over a stainless steel base, fit into the
two ends of a glass tube, which is periodically evacuated. The
absorption cell is 10.5 cm long and is housed inside a stainless
steel vacuum chamber, which is equipped with a pair of optical
windows of the same type as those of the cell. Consisting of a
cylindrical cavity inside an aluminium block, with inner and
external surfaces carefully polished, the cell is temperature
stabilized by means of four Peltier elements. Three precision
platinum resistance thermometers (Pt100) measure the temperature
of the cell's body while a proportional integral derivative
controller is used to keep the temperature uniform along the cell
and constant within 40 mK, over a time interval of $\sim$2 hours.
This active system also allows us to vary the gas temperature
between 270 and 330 K.

The Pt100 thermometers were calibrated at the triple point of
water and at the gallium melting point with an overall accuracy
better than 0.01 K. Both during the calibration and when placed in
the absorption cell, the thermometers are fed by a 1 mA current.
The correction due to the self heating effect has been applied
together with the associated uncertainty.

The sample cell is filled with CO$_{2}$ gas (with a nominal purity
of 99.999\%) at a pressure between 70 and 130 Pa, measured using a
1300 Pa full-scale capacitance gauge with a 0.25\% accuracy. A
turbomolecular pump is used to periodically evacuate the sample
cell and create high-purity conditions. Data acquisition is
performed by means of a digital oscilloscope (Tektronix TDS7104),
with a 13-bit vertical resolution (on the voltage range covered by
the absorption profile) and a total number of points for each
spectrum equal to 5000. Spectral averaging over 25 consecutive
scans is performed, the scan rate being 5 Hz. The resulting
spectrum is then transferred to a personal computer.
Simultaneously, the transmissions from both the optical resonator
and the etalon are recorded.

\begin{figure}
\centering
\includegraphics[bb=15 15 300 220, scale=.75]{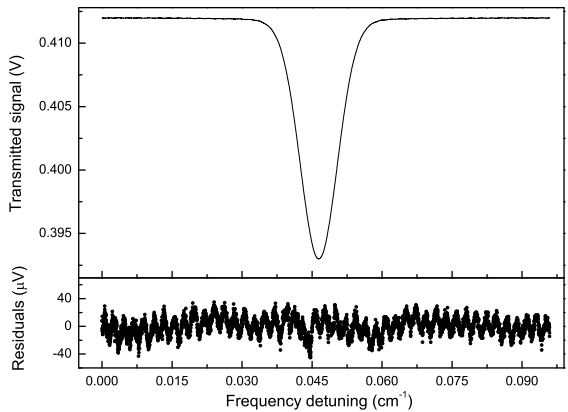}
\caption{\label{fig:4:two} Transmission spectrum for the R(12)
line at $\sim$100 Pa and 275.12 K. The absolute residuals are also
shown, as a result of a non-linear least-squares fit to a Voigt
convolution. The retrieved Doppler width was 268.01 MHz.}
\end{figure}


An example spectrum is shown in Fig.~\ref{fig:4:two}. The observed
S/N was 1600:1, with an equivalent noise detection bandwidth of
600 Hz and a power incident upon the cell of ~50 $\mu$W. A MATLAB
code is used to carry out the line shape fitting and retrieve the
Doppler width in a fully automated way for any set of repeated
acquisitions. Particular attention is paid to the frequency
calibration of the spectra. In fact, prior to each fit, the same
code analyzes the transmission spectrum from the confocal
resonator in order to construct a linear frequency scale that has
to be matched to the corresponding absorption spectrum. This is
done by means of a cubic spline interpolation of the peak
frequencies versus their index in the data array. Obviously, an
accurate determination of the cavity free-spectral-range,
$f_{\mathrm{FSR}}$, is also necessary. For that purpose, two
different methods were implemented, whose results were in total
agreement. The first one was based on the simultaneous detection
of absorption on both the zero- and first-order beams as provided
by the AOM, once the former was sent to a second absorption cell
(see Fig.~\ref{fig:4:one}). The two cells were filled with
CO$_{2}$ gas at the same pressure and at room temperature. Because
of the frequency shift between the two beams (80517.42$\pm$0.12
kHz, as measured by a universal counter), the center frequencies
of the two line profiles, simultaneously recorded along with the
transmission comb from the optical cavity, are separated by an
exactly known quantity. The second method consisted in making a
broad laser frequency scan ($\sim$0.5 cm$^{-1}$) in order to
observe a pair of CO$_{2}$ absorption lines, namely the R(8)
$\nu_{1}+2\nu_{2}^{\phantom{1}0}+\nu_{3}$ and R(27)
$\nu_{1}+3\nu_{2}^{\phantom{1}1}-\nu_{2}^{\phantom{1}1}+\nu_{3}$
transitions, whose center frequencies are accurately known. These
frequencies, as well as that of the R(12) line, were taken from
Ref. \onlinecite{miller-2004-4}. Thus, as a result of a weighted
mean between the two values, $f_{\mathrm{FSR}}$ was found to be
149.56$\pm$0.01 MHz. In our operation conditions, the effects of
temperature variations and changes in the cavity alignment on
$f_{\mathrm{FSR}}$ could be neglected. As it will be shown later
on, the relative uncertainty on the cavity free-spectral-range,
multiplied by a factor of 2, will contribute to the overall
accuracy of the spectroscopic determination of $k_{\mathrm{B}}$.
The linearity of the frequency scale was periodically checked
using the transmission spectra from the germanium etalon.
Deviations from linearity, which could affect the symmetry of the
absorption profiles, were found to be negligible at the present
experimental accuracy. Coming back to the fit of the absorption
line shape, the normalized Voigt function was computed using the
approximation of Weideman \cite{Weideman-1994-4}. It was confirmed
that this approximation is equal to the Voigt convolution, as
tabulated in Ref. \onlinecite{abramowitz-1972-4}, to better than 1
part in 10$^{6}$. The nonlinear least-squares fitting code was
based on a Levenberg-Marquardt algorithm. The code was tested on
simulated spectra, assuming a gas pressure of 130 Pa at a
temperature of 296 K, and adding a random noise to the computed
line shape. As one could expect, the relative accuracy in the
retrieval of the Doppler width strongly depends on the
signal-to-noise ratio. In fact, relative deviations between the
real value and the retrieved one of 2.6$\times$10$^{-5}$ and
5.2$\times$10$^{-6}$ (mean values over 10 simulations) were found
for S/N=1600 and 1.6$\times$10$^{4}$, respectively.

\begin{figure}
\centering
\includegraphics[bb=15 16 306 217, scale=0.75]{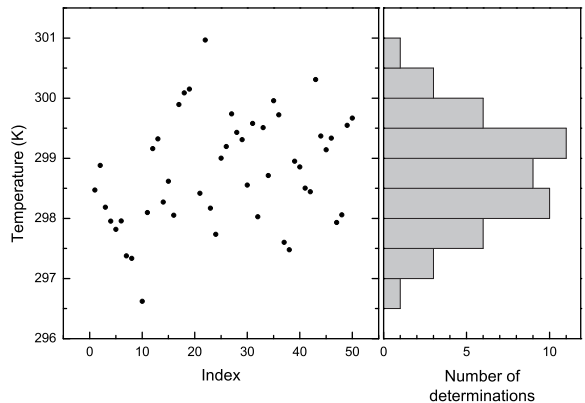}
\caption{\label{fig:4:three} Repeated spectroscopic measurements
of the absolute temperature and a histogram of the determinations.
The mean value is 298.77 K, the standard deviation 0.89 K, and the
standard error 0.13 K.}
\end{figure}

The exponential of a Voigt function is able to fit our measured
spectra within the experimental noise, as shown in
Fig.~\ref{fig:4:two}. Reported in the bottom part of the figure,
the residuals evidence that the noise is dominated by periodic
fluctuations due to spurious etalon effects, whose
root-mean-square value is of the order 10$^{-5}$ V. Owing to our
intensity stabilization procedure, the background signal for the
Voigt fit could be optimally reproduced by simply adding a flat
baseline to the line shape function. Apart from the Doppler line
width, the remaining free parameters were the line center
frequency, the homogeneous width and an amplitude factor, the
latter representing the product of the line strength factor, the
gas pressure and the absorption path length. In our experimental
configuration, saturation and transit time broadening could be
neglected. It is worth noting that the scan goes quite far into
the line wings, as demonstrated \textit{a posteriori} by the area
under the normalized fitting function, which was found to be
larger than 0.99.

In Fig.~\ref{fig:4:three}, an example of spectroscopic
determinations of the gas temperature, over a time interval of
$\sim$4 minutes, is reported. The histogram, on the right side of
Fig.~\ref{fig:4:three}, essentially shows that the distribution is
gaussian. Simultaneously to the spectral acquisitions, the Pt100
temperatures were also recorded, providing a mean value of
298.68~K, with a standard deviation of $\sim$10~mK. With very few
exceptions, a good agreement was found between the spectroscopic
and Pt100 temperatures, typical deviations being much smaller than
1~K.

\begin{figure}
\centering
\includegraphics[bb=0 20 310 230, scale=.75]{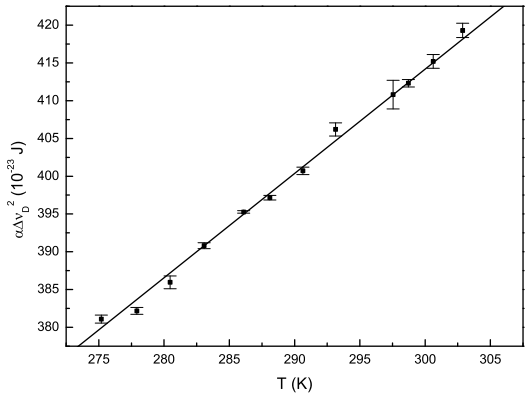}
\caption{\label{fig:4:four} A plot of the quantity
$\alpha\Delta\nu_{D}^{2}$ as a function of the gas temperature,
where $\alpha=mc^{2}/(8\nu^{2}\ln2)$ and $\alpha\Delta\nu_{D}^{2}$
represents the thermal energy. The slope of the weighted best-fit
line directly provides the value of $k_{\mathrm{B}}$. Each point
represents the mean value over a number of repeated
determinations, the error bars being the standard error
(3$\sigma$). These data resulted from more than 1500 spectra,
acquired over five consecutive weeks.}
\end{figure}

Doppler width determinations were repeated as a function of the
gas temperature, in the range between the triple point of water
and the gallium melting point. This approach was mainly adopted to
cancel out possible systematic errors arising from residual etalon
effects due to the cell's windows. In Fig.~\ref{fig:4:four}, the
experimental values of the quantity
$\frac{m}{8\ln2}\frac{c^{2}}{\nu^{2}}\Delta\nu_{D}^{2}$ and the
corresponding temperatures, as provided by the Pt100 thermometers,
are plotted. The molecular mass was calculated from the relative
atomic masses of $^{16}$O and $^{12}$C, the molar mass constant,
and the Avogadro number \cite{mohr-2005-4}. A linear fit of these
data directly provides the spectroscopic value for the Boltzmann
constant, namely $k_{\mathrm{B}} = 1.380 58 (22) \times 10^{-23}$
\textrm{JK$^{-1}$}. The relative uncertainty in the temperature
values, also including the short-term fluctuations, were found to
be a factor of 10 smaller compared to the relative error of a
typical spectroscopic determination of the temperature itself. For
this reason, there errors were ignored in the linear fit. The
1-$\sigma$ uncertainty on $k_{\mathrm{B}}$ includes the
statistical error on the slope of the best-fit line as well as the
uncertainty on the frequency scale. Our determination agrees with
the CODATA value.


In conclusion, we have demonstrated that highly accurate,
intensity-stabilized, laser absorption spectrometry in the
near-infrared represents an extremely powerful tool for primary
gas thermometry. Our experiment allowed us to provide a
spectroscopic measurement of the Boltzmann constant, with a
relative accuracy of 1.6$\times$10$^{-4}$. The absorption features
of our molecular target around 2-$\mu$m were found to be a very
good choice mainly because of the insensitivity of the Doppler
width to the gas pressure. More precisely, for CO$_{2}$ pressures
below a few hundred Pa, we did not observe any influence from
Dicke narrowing or speed-dependent effects. This is true at the
present precision level. Once, in the near future, we will be able
to retrieve the Doppler width with a much smaller uncertainty, it
will be necessary to confirm its behavior as a function of the gas
pressure.

Future developments include the use of a pair of phase-locked
extended-cavity diode lasers to improve our capability of
measuring laser frequency variations, as well as the
implementation of a new detection method based on fast amplitude
modulation of the laser beam and phase-sensitive detection to
enhance the signal-to-noise ratio. We are also building a new
temperature-stabilized cell, which should ensure a constant and
homogeneous gas temperature within 1 mK. After these significant
improvements, our method could contribute to the new definition of
the kelvin, in the near future. In fact, we will no longer be
limited by the accuracy of the frequency scale. On the other hand,
the S/N is expected to increase by a factor of 25, provided that
spurious etalon effects are eliminated. We also plan to exploit
the excellent linearity of standard InGaAs detectors by shifting
the operation wavelength down to 1.4 $\mu$m. According to our
simulations, it should be possible to reach an accuracy level of a
few parts in 10$^{6}$.

\begin{acknowledgments}
This work is supported by the Italian Ministry for University and
Research, under the framework program PRIN 2006. The authors would
like to thank  Massimo Inguscio for a critical reading of the
manuscript and Guglielmo Tino for seminal discussion.
\end{acknowledgments}


\end{document}